\documentclass[10pt,twocolumn,letterpaper]{article}

\usepackage{iccv}
\usepackage{times}
\usepackage{epsfig}
\usepackage{graphicx}
\usepackage{amsmath}
\usepackage{amssymb}
\usepackage{multirow}
\usepackage{adjustbox}
\usepackage{booktabs}
\usepackage{tabularx}
\usepackage{bm}
\usepackage[table,xcdraw,dvipsnames]{xcolor}
\usepackage{color, colortbl}
\usepackage{authblk}

\definecolor{Gray}{gray}{0.9}
\definecolor{brightturquoise}{rgb}{0.85, 1, 1}
\newcommand{\mycc}{\cellcolor[HTML]{EDF6FF}}

\newcommand{\reffig}[1]{\text{Figure~\ref{#1}}}
\newcommand{\reftab}[1]{\text{Table~\ref{#1}}}
\newcommand{\refeq}[1]{\text{Eq.~\ref{#1}}}

\usepackage[pagebackref=true,breaklinks=true,colorlinks,bookmarks=false,citecolor=blue,linkcolor=red]{hyperref}

\iccvfinalcopy 

\def\iccvPaperID{11495} 

\ificcvfinal\fi

\begin{document}

\title{


Lip2Vec: Efficient and Robust Visual Speech Recognition via Latent-to-Latent Visual to Audio Representation Mapping
}

\author[1,2]{Yasser Abdelaziz Dahou Djilali}
\author[1]{Sanath Narayan}
\author[1]{Haithem Boussaid}
\author[1]{Ebtessam Almazrouei}
\author[1]{Merouane Debbah}
\affil[1]{Technology Innovation Insitute, UAE}
\affil[2]{Dublin City University, Ireland}

\maketitle
\ificcvfinal\thispagestyle{empty}\fi

\begin{abstract}


Visual Speech Recognition (VSR) differs from the common perception tasks as it requires deeper reasoning over the video sequence, even by human experts. Despite the recent advances in VSR, current approaches rely on labeled data to fully train or finetune their models predicting the target speech. This hinders their ability to generalize well beyond the training set and leads to performance degeneration under out-of-distribution challenging scenarios. Unlike previous works that involve auxiliary losses or complex training procedures and architectures, we propose a simple approach, named Lip2Vec that is based on learning a prior model. Given a robust visual speech encoder, this network maps the encoded latent representations of the lip sequence to their corresponding latents from the audio pair, which are sufficiently invariant for effective text decoding. The generated audio representation is then decoded to text using an off-the-shelf Audio Speech Recognition (ASR) model. The proposed model compares favorably with fully-supervised learning methods on the LRS3 dataset achieving 26 WER. Unlike SoTA approaches, our model keeps a reasonable performance on the VoxCeleb test set. We believe that reprogramming the VSR as an ASR task narrows the performance gap between the two and paves the way for more flexible formulations of lip reading. 


\end{abstract}

\section{Introduction}

The process of inferring visual cues from a speaker's facial expressions and lip movements to interpret speech in a silent setting is refereed to as lip-reading or visual speech recognition (VSR). VSR is mostly useful in environments where the speech is unclear or difficult to hear due to some confounding factors \cite{bourguignon2020lip, bernstein2000speech}. Hearing and speech-impaired individuals also greatly benefit from VSR \cite{Tye-Murray2007AudiovisualHearing}. Albeit the small variations around the mouth area, the space of spoken words can be large due to the phonemes composition mechanism. This makes the task highly ambiguous as several phonemes incur similar visual characteristics. Moreover, VSR needs to be robust to variations \wrt multiple speakers, head pose movements, non-verbal facial expressions and imaging conditions. Furthermore, lip-reading requires the integration of visual features and contextual information (\ie, topic, key words search, environment and place, \etc) \cite{sumby1954visual, massaro1990perception, bernstein2022lipreading}.
Over the last few years, computational methods for VSR has seen a surge with the recent proposed datasets, and can be grouped into (\textit{i}) word-level prediction that classifies a silent video segment into a pre-defined vocabulary of words; (\textit{ii}) continuous visual speech recognition, which predicts sentences for varying length video sequences.

Most existing VSR approaches employ a common pipeline, where lip sequences are spatially encoded using a convolution-based backbone and passed to a contextual encoder (\ie, transformer \cite{vaswani2017attention} or conformer \cite{gulati2020conformer}) to model temporal dependencies. Finally, auto-regressive transformer decoder cross-attends to these representations for predicting the text. Previous works focused on enhancing the video representations for better decoding, while early approaches pretrained the backbone on word-level LRW dataset \cite{chung2017lip} for better convergence on continuous VSR \cite{afouras2018deep, ma2021lira}. In contrast, \cite{ma2022visual,Afouras2019ASRReading} exploit audio information as an extra supervision for an auxiliary task. Recently, cross-modal self-supervised pretraining has been a dominant paradigm for a smoother supervised finetuning afterwards \cite{shi2022learning,shi2022robust, haliassos2022jointly}.


Alternatively, the audio latent space exhibits the properties of local smoothness between input and its representation, is temporally coherent over a sequence of observations, has simple dependencies among its factors and is sparsely activated for a specific input, leading to robust and performing models \cite{baevski2020wav2vec, baevski2019vq, radford2022robust, roger2022deep}. Whereas the lip sequence is more ambiguous, with complex dependencies over the sequences as the movements are only a partial observation of a larger system that includes tongue, and other facial muscles\cite{fitch2000evolution}. Thus, this highlights a fundamental question about supervised learning on lip-reading data that is likely to result in local generalization, while lacking robustness on out-of-distribution data. In this work, we study these questions, uncovering key representational analogies between audio and lip sequences, the ways in which these analogies can act as a robust support for downstream task transfer, allowing for reprogramming the VSR using off-the-shelf ASR models. Specifically, our contributions are:

\begin{itemize}
    \item We propose Lip2Vec framework that simulates VSR as an ASR task by learning a prior network that maps lip sequence features to audio-like representations, which can then be decoded to text using an ASR model.

    \item Through extensive evaluation, we show that learning the prior network can be exploited for decoding text. Furthermore, it performs on par with fully supervised methods on the LRS3~\cite{afouras2018lrs3} test set and generalizes better on the VoxCeleb2-en~\cite{chung2018voxceleb2} test set.
    
    \item Our approach addresses the generalization and robustness challenges encountered by VSR models. The design explicitly bridges the gap between the VSR and ASR performances, that is proportional to the quality of the learned prior network.

    \item Our approach benefits from CTC-only decoding of ASR models and is 10$\times$ faster compared to standard VSR approaches, which decode text auto-regressively.

\end{itemize}

\section{Related Works}
Here, we briefly discuss the works related to the task of visual speech recognition.

\subsection{Visual Speech Recognition}

Sentence-level VSR, also referred as continuous visual speech recognition is challenging due to unconstrained large corpus and complex dependencies across the sequence length with regards to the text target.  Whilst we briefly overview the recent sentence-level VSR efforts, we refer to \cite{Sheng2022DeepSurvey,Zhou2014ADecoding,Fenghour2021DeepSurvey} for extensive reviews. Learning from scratch on VSR datasets \cite{afouras2018lrs3, afouras2018deep} raises serious optimization issues. This difficulty emerges as the decoder cross-attention is under-optimized in early training, resulting in noisy contextual information for the queries. 

Several hypotheses have been proposed to account for this. The work of \cite{ma2021end} proposed a curriculum learning approach, where shorter sequences are initially used for training followed by progressively adding longer ones. Differently, VTP \cite{prajwal2022sub} proposed sub-words learning scheme using frame-word boundaries to crop out training samples for a better convergence. These training strategies are computationally demanding and hard to scale to larger datasets. The recent works of \cite{ma2022visual, Afouras2019ASRReading} proposed to exploit the audio latent representations as part of a auxiliary task, where the network is optimized to predict pretrained ASR representations along with the target text, making the optimization more stable as it provides extra supervision. Intuitively, if the transformer encoder is able to match the audio features statistics, it has to adjust the attention weights for a better decoding. Another line of research leverages pretraining on larger datasets in a self-supervised way (SSL), then finetuning on labeled VSR data using video-text pairs \cite{shi2022learning,shi2022robust,haliassos2022jointly,zhu2022vatlm, ma2021lira}. AV-HuBERT \cite{shi2022learning} fuses the masked audio-visual representations to predict the cluster assignments created from the audio features, thus, distilling knowledge from the audio stream features to model visual inputs. VATLM \cite{zhu2022vatlm} attempts unifying both modalities using a one tower design, where a single network is optimized to construct a common representation space for video, audio and text. This is achieved by setting a unified tokenizer for all modalities, and then performing the masked prediction task over the unified tokens. The works of \cite{Sheng2021Cross-modalTraining,Ma2021ContrastiveRepresentations} designed cross-modal self-supervised learning frameworks, by adopting contrastive learning \cite{jaiswal2020survey} to learn discriminative visual representations that appear to improve VSR performance and generalization. Recently, RAVen \cite{haliassos2022jointly} designed an asymmetric SSL framework to induce a one-way knowledge distillation, where the audio networks predict both audio and video representations, whereas the visual network is restricted to predict the audio features only. This forces the audio network to serve as a strong teacher, as it would adjust to both modalities at the same time.

In this work, we argue that, despite the remarkable results of SSL pretraining, its expressive power can be further exploited differently. One design choice is to freeze the learned representation for the downstream task of VSR. Unlike the classification setting, the common practice of linear probing \cite{chen2020simple} is not effective on the VSR datasets \cite{ma2021lira}. The contributions of this paper attempt to address this question.

\subsection{Latent-to-Latent Models}
Over the last few years, latent-to-latent approaches have attracted much attention especially in the cross-modal generation literature. The high-level idea aims to match representations from two  manifolds unified by a unique generating process, where correspondences are recovered and knowledge from one domain is transferred to another. In fact, Dall-e1\cite{ramesh2021zero} trained a prior network on large scale datasets to map text to image tokens so as to perform text-guided image generation using VQ-VAEs \cite{van2017neural}, while in the work of \cite{Zhu2022QuantizedVideos}, latent-to-latent network is employed to map dense visual features to discrete music representation. The work of
\cite{Khodadadeh2022LatentImages} manipulates the GAN's latent space by steering the representations to change facial attributes.  Adversarial reprogramming \cite{neekhara2022cross} was taken out of the realm of adversarial attacks to repurpose an image classification to perform sequence classification tasks \cite{neekhara2022cross}.  In this work, we take a step forward and extend latent-to-latent techniques to VSR task, which is more fine-grained and requires better temporal modeling.

\section{Method}


As mentioned earlier, audio encoders of ASR models learn to transform the audio inputs to well-structured latent representations that are sufficiently robust for the task of text decoding. Our approach takes advantage of these audio representations by utilizing them as targets for training a differentiable  parametric function $\bm{f}_{\theta}: \mathbf{z_{v}} \mapsto \boldsymbol{z_{asr}}$, with parameters $\theta$ (\eg, a neural network). Such a prior network transforms video latent representations computed by a video encoder to synthetic audio representations, which are then input to the corresponding ASR decoder for predicting the text. Our prior network is optimized to model the joint distribution over the video and audio representations by maximizing the cosine similarity between the respective representations of the pairs.

\subsection{Preliminaries}


We call for a function $\boldsymbol{f_{\omega}}: \boldsymbol{V}^{T \times W \times H} \mapsto z_{\mathbf{v}}$. This function is trained in a self-supervised way such that it encodes the lip sequences by explicitly capturing the characteristics of the lip motion (\ie, temporal smoothness, invariances of small and local changes in the lip sequences, \etc), while still being unconditioned by the text labels. For the audio modality, the goal is to learn a model  $\boldsymbol{f_{\gamma}}: \boldsymbol{A}^{T \times S} \mapsto y$, which maps the input audio signal to the corresponding text labels $y$.


\begin{figure}[t]
\makebox[\linewidth]{
    \centering
    \includegraphics[width=1.00\columnwidth]{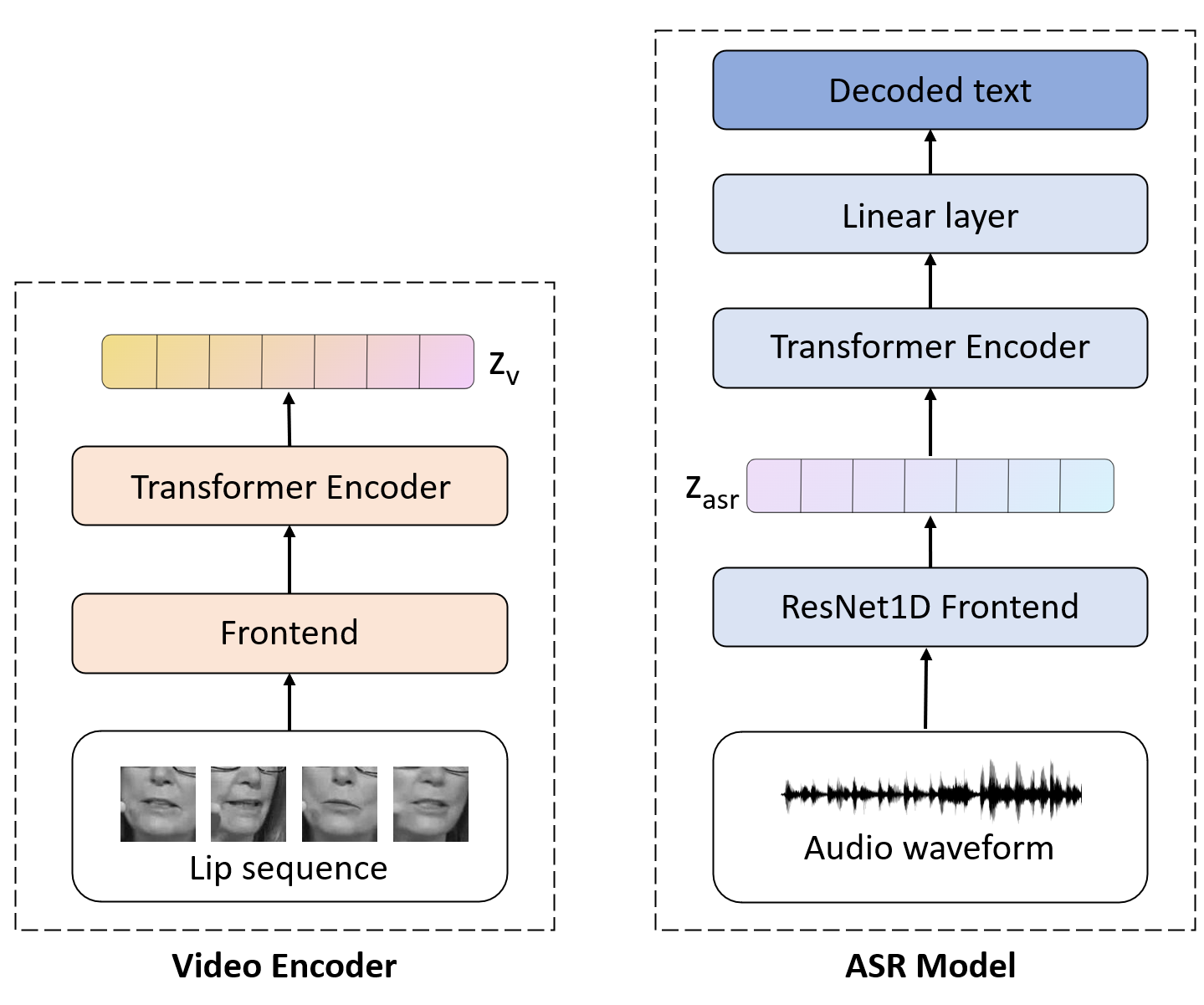}}
    \caption{\textbf{On the left:} The video encoder takes a sequence of frames as input and computes the corresponding video representation $z_{\mathbf{v}}$. \textbf{On the right:} The frontend of the ASR model takes an audio input and obtains the audio representation $z_{\mathbf{asr}}$, which is the passed through a transformer encoder and linear layer for obtaining the text output.}
    \label{fig:encoders}
\end{figure}

\noindent\textbf{Video encoder:} We adopt the self-supervised learned model from AV-HuBERT~\cite{shi2022learning} as our video encoder. It comprises a modified ResNet~\cite{ResNet} as frontend, followed by a transformer encoder. The 2D frontend of ResNet is replaced with a 3D convolutional layer~\cite{petridis2018audio}. The AV-HuBERT model is pretrained to solve the masked-prediction task, where given the masked audio and video representations output by the ResNet, the goal is to predict the clusters assigned using acoustic features (\eg, MFCC). This is iteratively refined using k-means clustering of features learned by the audio-visual encoder. Consequently, the encoder learns to better encode the characteristics of a video sequence. Given a video sequence in $\mathbb{R}^{T \times W \times H}$, the video encoder $\boldsymbol{f_{\omega}}(\cdot)$ maps it to  $z_{\mathbf{v}} \in \mathbb{R}^{T \times D}$. \reffig{fig:encoders} (left) shows the video encoder architecture for extracting $\mathbf{z_{v}}$ from a video sequence.

\noindent\textbf{ASR model:} While our framework can host any off-the-shelf ASR model, we leverage Wav2Vec2.0 \cite{baevski2020wav2vec} for its simplicity and generalization capacity.
Its contrastive pretraining maximizes the mutual information between a set of anchors from contextualized representations output by the transformer encoder, and their positive pair samples from quantized representations of the ResNet features, while pushing away the set of negatives. Such a pretraining on $53$k hours of unlabeled data promotes better temporal modeling and achieves a low WER of 4.8 on Librispeech~\cite{panayotov2015librispeech} even when finetuning on just ten minutes of labeled data. The ASR model $\boldsymbol{f_{\gamma}}(\cdot)$ maps an acoustic signal to audio representations $\boldsymbol{z_{asr}}$ using a feature extractor and projector. The $\boldsymbol{z_{asr}}$ is then contextualized by the transformer encoder and mapped to a vocabulary of 32 characters using a linear layer, making it faster compared to auto-regressive decoding techniques \cite{synnaeve2019end}. \reffig{fig:encoders} (right) shows the pipeline for decoding the text from an audio input. 


\subsection{Learning the Prior Network}

We freeze the encoders ($\boldsymbol{f_{\gamma}}(\cdot)$ and $\boldsymbol{f_{\omega}}(\cdot)$) and learn the prior distribution over video and audio latents by maximizing the cosine similarity between $\boldsymbol{z_{v}}$ and $\boldsymbol{z_{asr}}$ \wrt  $\boldsymbol{p_{\theta}}$ \cite{ramesh2021zero}. To this end, we instantiate the prior network $\boldsymbol{f_{\theta}}(\cdot)$ as a standard transformer encoder \cite{vaswani2017attention}.


Given a video-audio pair as inputs, we employ $\boldsymbol{f_{\omega}}(\cdot)$ to encode the lip sequence, whereas the audio signal is encoded with $\boldsymbol{f_{\gamma}}(\cdot)$ up to the ResNet level. The video representations $\mathbf{z_{v}}$ are summed with their corresponding masked audio features $\mathcal{M}(\mathbf{z_{asr}})$, where $\mathcal{M}(\cdot)$ denotes the time-masking operation. The resulting combined representation is modeled as a single data source to generate the synthetic audio representations $\mathbf{z_{asr}^g}$. The prior network is an encoder-only transformer model that exploits the expressive power of the self-attention to perform the manifold mapping from video to audio. Moreover, the task is to model the joint distribution over video and audio representations by associating the recurring patterns, compare their dependencies, and infer analogies on how the lip movements can be synthesized as an audio signal. Finally, the prior network $\bm{f}_\theta(\cdot)$ is optimized to predict the unmasked audio representations.

\begin{figure}[t]
\makebox[\linewidth]{
    \centering
    \includegraphics[width=1.00\columnwidth]{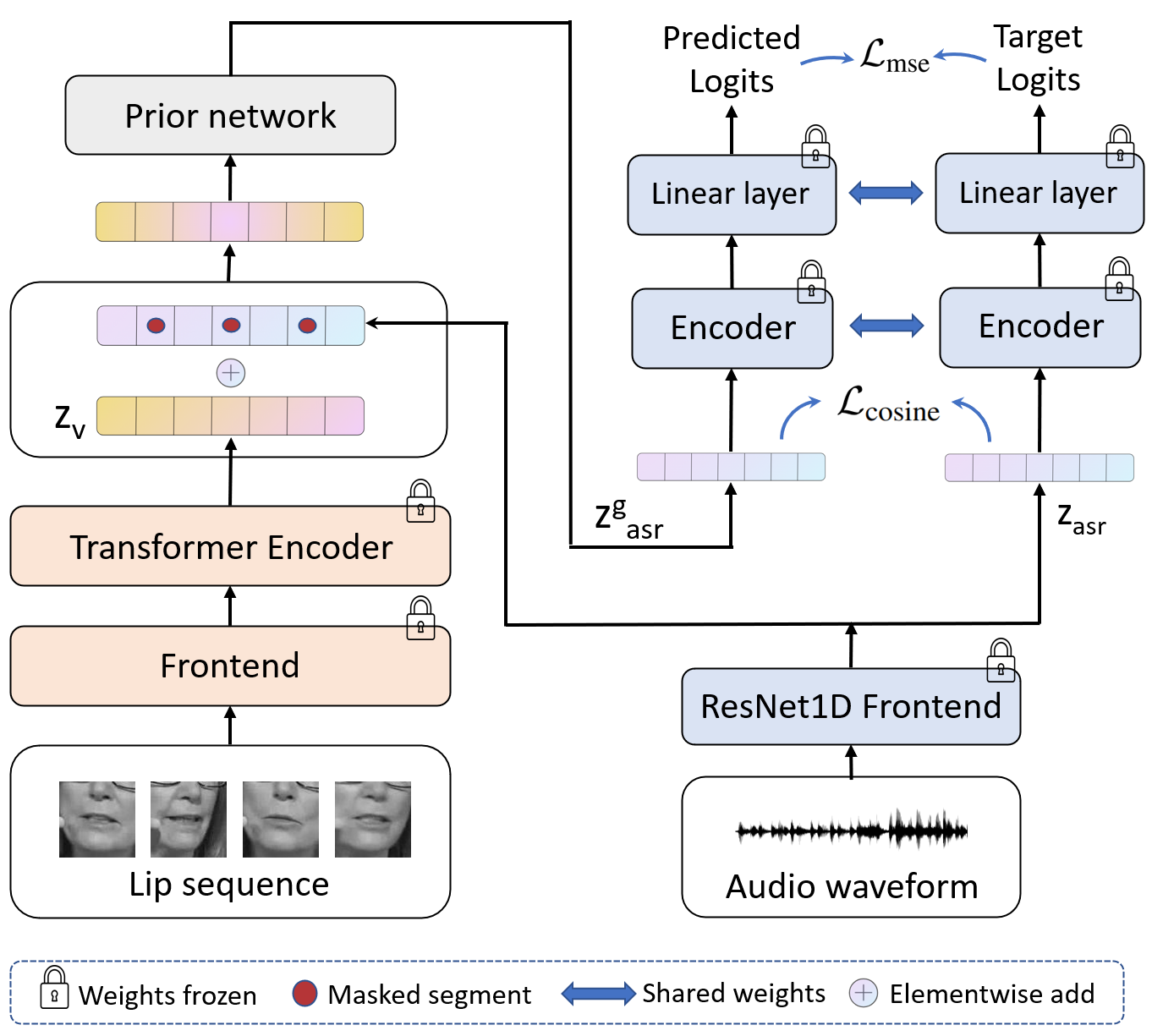}}
    \caption{\textbf{Training pipeline of our Lip2Vec framework.} The video representations $\mathbf{z_{v}}$ are summed with the masked audio representations $\mathcal{M}(\mathbf{z_{asr}})$ and input to the prior network. The prior network generates corresponding synthetic audio representations $\mathbf{z_{asr}^g}$, which are compared with the original $\mathbf{z_{asr}}$ through a cosine similarity loss ($\mathcal{L}_{cosine}$). Furthermore, the representations $\mathbf{z_{asr}^g}$ and $\mathbf{z_{asr}}$ are passed independently through the transformer encoder and linear layer of the ASR model to obtain the predicted and target logits, respectively and aligned through an MSE loss ($\mathcal{L}_{mse}$). Note that the video encoder and the ASR model parameters are kept frozen throughout the training.\vspace{-0.2cm}}
    \label{fig:training}
\end{figure}

\noindent\textbf{Avoiding collapse:} Albeit representing the same target speech, the audio and video manifolds are likely disjoint and are may not transport easily. 
In the process of maximizing the similarity between the respective representations, the task is to construct an input stream that achieves the optimization sweet-spot, thereby allowing the prior network to smoothly learn the mapping between the two manifolds. Furthermore, utilizing only video representations as input leads to degraded performance due to the difficulty in optimization resulting from missing informative features. In contrast, utilizing the audio representations summed with the video representations results in the prior network relying solely on the former for the prediction, while neglecting the latter completely and results in degraded VSR performance. To alleviate these issues of collapse, we opt for a masking schedule over the audio representations, where the mask proportion is linearly increased as the training progresses and ensures the input stream is video features only during the final epochs of training. Such a progressive masking of audio representations at the input of the prior network promotes smoother optimization during the early stage of training, and pushes the transformer to slowly learn the generalizable features for the VSR task.

\begin{figure}[t]
\makebox[\linewidth]{
    \centering
    \includegraphics[width=0.9\columnwidth]{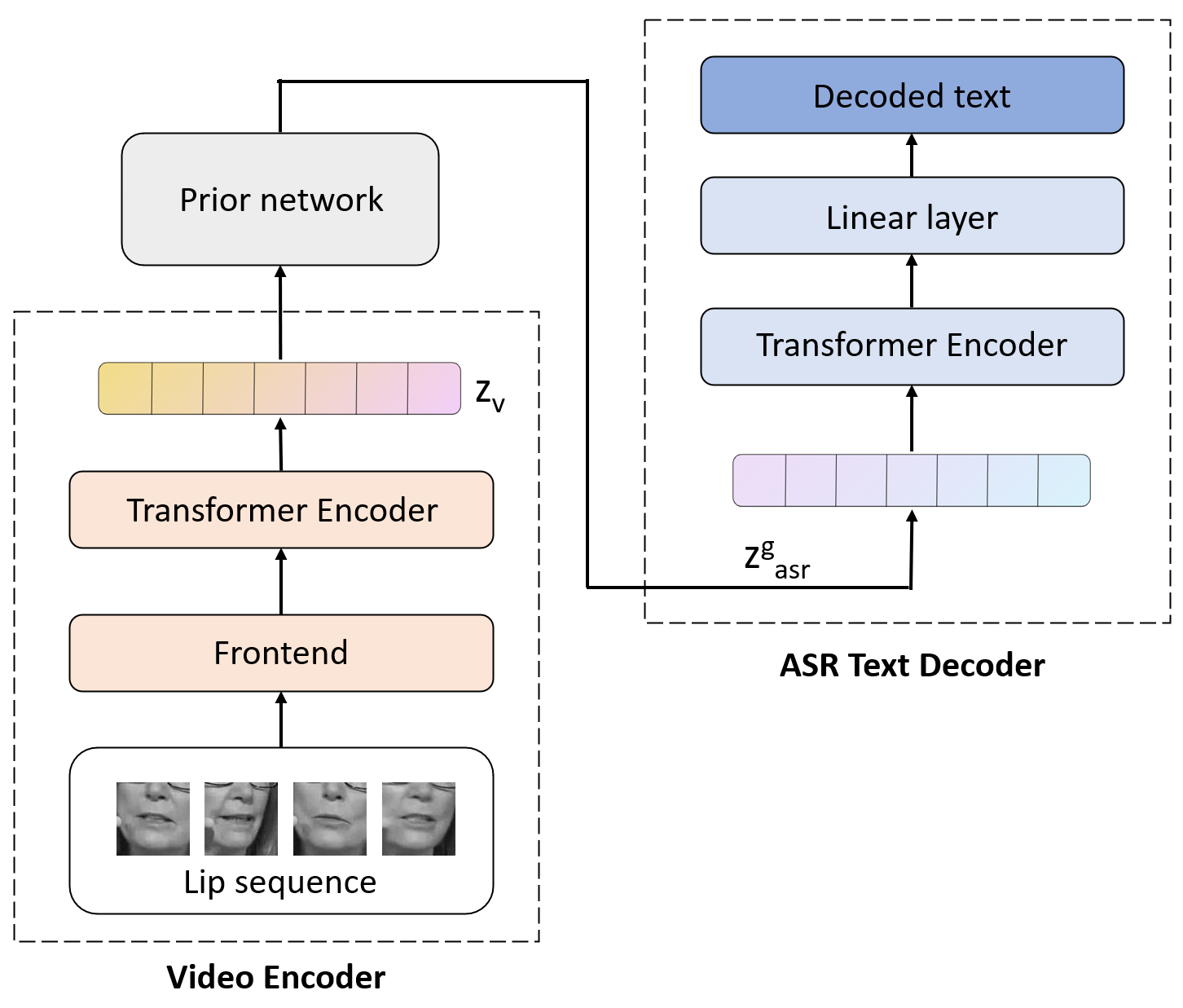}}
    \caption{\textbf{Decoding text from video during inference.} The video representations $\mathbf{z_{v}}$ computed by the video encoder are input to our learned prior network, which synthesizes audio representations $\mathbf{z_{asr}^g}$. These representations are then passed through the encoder and linear layer of the ASR model to predict the text. Note that audio representations are not used at test time.}
    \label{fig:inference}
\end{figure}

\subsection{Training and Inference}


\noindent\textbf{Training:} For a pair of video and audio representations $\mathbf{z_{v}}$ and $\mathbf{z_{asr}}$, the prior network optimizes:
\[ f_{\theta}: \mathbf{z_{in}} \mapsto \mathbf{z_{asr}^{g}}, \text{where}  \quad  \mathbf{z_{in}} = \mathbf{z_{v}} + \mathcal{M}(\mathbf{z_{asr}}).\]
We define $\mathcal{M}(\cdot)$ as the masking operation with a probability $p$ that is a function of training steps. 
%
Given the audio input, the corresponding representations $\mathbf{z_{asr}}$ and logits $\mathbf{h_{asr}} \in \mathbb{R}^{T\times C}$ (with $C$ being the vocabulary size of the ASR model) are utilized as targets for optimization. Here, $\mathbf{z_{asr}}$ is extracted at the ResNet level, while the logits are extracted after the final linear projection layer of the Wav2Vec2.0. The training objective is to minimize the negative cosine similarity between representations summed over the temporal dimension, while maintaining a small distance with logits. Particularly, the losses are given by

\begin{equation}\label{eq1}
    \begin{aligned}
    \mathcal{L}_{\text{cosine}} = - \sum_{i=1}^{T} \mathbf{z}_{\mathbf{asr}, i} ^\top \mathbf{z}_{\mathbf{asr}, i}^{\mathbf{g}}, \quad \text{and}
    \end{aligned}
\end{equation}

\begin{equation}\label{eq2}
    \begin{aligned}
    \mathcal{L}_{\text{mse}} = \frac{1}{T} \sum_{i=1}^{T} (\mathbf{h}_{\mathbf{asr}, i}^{\mathbf{g}} - \mathbf{h}_{\mathbf{asr}, i})^{2}.
    \end{aligned}
\end{equation}
The final objective function is given by
\begin{equation}\label{eq3}
    \begin{aligned}
    \mathcal{L} = \mathcal{L}_{\text{cosine}} + \alpha \mathcal{L}_{\text{mse}},
    \end{aligned}
\end{equation}
where $\alpha$ is a hyperpameter for weighting the MSE loss. 

\noindent\textbf{Inference:} At test time, a query video is input to the video encoder to obtain the video representation $\mathbf{z_{v}}$. The prior network takes this $\mathbf{z_{v}}$ and generates a corresponding audio representation $\mathbf{z^g_{asr}}$, which is then passed to the transformer encoder and linear layer of the ASR model to obtain the predicted text $\hat{y}$. \reffig{fig:inference} shows our inference pipeline for decoding the text from a video-only input. Note that audio is not utilized at inference time for decoding the text.

\section{Experiments\label{sec:exp}}

\noindent\textbf{Datasets:} We train the prior network using the video-audio pairs on: LRS3~\cite{afouras2018lrs3} and VoxCeleb2-en~\cite{chung2018voxceleb2}. The LRS3 dataset comprises a total of $433$ hours of training videos from pretrain and trainval sets. From the multi-lingual VoxCeleb2 dataset, a subset of $1326$ hours of videos for the English language is selected as VoxCeleb2-en, as in~\cite{shi2022learning}. We evaluate the prior network using two test sets of LRS3 and VoxCeleb2-en, as detailed below:


\begin{itemize}
    \item LRS3: a small scale test set of around $1$ hour in total, consisting of $1321$ sequences. We leverage the $68$ facial landmarks provided by~\cite{ma2022visual} to crop the utterances around the mouth area.
    
    \item  VoxCeleb-En: we randomly sample $5$K videos from the VoxCeleb2-en test set, with the same duration statistics as the LRS3 test set. We use Whisper medium \cite{radford2022robust} as the labeling tool to obtain the text transcripts. Moreover, for efficiency reasons, we utilize Yolo5Face \cite{qi2023yolo5face} to obtain the landmarks instead of relying on RetinaFace \cite{deng2020retinaface, bulat2017far} face detector. We found that the resulting $5$ facial landmarks are sufficient for cropping the mouth regions \footnote{This new pseudo labelled test set test set will be made publicly available to serve as an extra benchmark for the community}. 
    
\end{itemize}

\noindent\textbf{Evaluation metric:} As in~\cite{ma2022visual}, we employ the word error rate (WER) to measure the matching between the predicted text and the ground truth transcript.


\noindent\textbf{Implementation details:}
We adopt the implementations of AV-HuBERT \cite{shi2022learning} and Wav2Vec2.0 \cite{baevski2020wav2vec} from the official fairseq repository\footnote{\href{Github}{https://github.com/facebookresearch/fairseq/tree/main/fairseq}}. For the prior network, we consider two configurations: BASE with $6$ transformer layers and LARGE with $12$ layers. The embedding dimension/feed-forward dimension/attention heads in each transformer layer are $768$/$3072$/$12$ for both variants. Furthermore, we employ a fully-connected layer and a temporal convolution upsampling to match the $50$ fps of the audio representations. Base and large are trained on the low and high ressource settings respectively. The prior network is implemented in PyTorch~\cite{pytorch} and trained using $4$ and $8$ NVidia A100 40GB GPUs for base and large models, respectively. All the models are trained for $30$ epochs using the AdamW~\cite{adamw} optimizer. We employ a warmup of $5$ epochs and a cosine learning rate scheduler with maximum $lr$ set to $10^{-3}$.

\noindent\textbf{On using labeled video-text data:} It is worth mentioning that the prior network weights are not fine-tuned using labeled data containing video-text pairs. Both video encoder and ASR models are kept frozen when performing the latent to latent training. The main motivation is to set a robust evaluation procedure and to prevent the prior network from adapting its parameters to represent the video as an audio, but rather to semantically match their latent spaces.

\subsection{Main Results}

\begin{table}[t]

\centering 
\caption{\textbf{Supervised finetuning \vs latent-to-latent training.} Comparison in terms of WER on LRS3 test set is shown. The same pretrained video encoder from AV-HuBERT~\cite{shi2022learning} is finetuned or utilized for the prior network. For the supervised learning, AV-HuBERT is trained with either linear layer (CTC) or a decoder (CE). Our Lip2Vec consistently improves the performance across different settings with simple CTC decoding.}

\setlength{\tabcolsep}{3pt}
\adjustbox{width=1\columnwidth}{
\begin{tabular}{lcc|ccc}
\toprule[0.1em]
 \multirow{2}{*}{\textbf{Encoder}} & \multirow{2}{*}{\textbf{Pretrain}} & \multicolumn{1}{c|}{\multirow{2}{*}{\textbf{Finetune}}}& \multicolumn{2}{c}{\textbf{Supervised}~\cite{shi2022learning}} & \textbf{Ours: Lip2Vec} \\
 &  &  & CTC & CE  & CTC     \\
 \toprule[0.1em]
\multirow{4}{*}{\textbf{Base}} & \multirow{2}{*}{433h} & 30h & 55.3 & 51.8  & 49.5 \\
 &  & 433h & 49.3 & 44.0  & 42.0 \\ \cmidrule{2-6}
 & \multirow{2}{*}{1759h} & 30h & 47.3 & 46.1 & 40.6 \\
 &  & 433h & 43.0 & 34.8   & 34.1 \\
 \midrule
\multirow{4}{*}{\textbf{Large}} & \multirow{2}{*}{433h} & 30h & 48.4 & 44.8 &  55.4\\
 &  & 433h & 44.3 & 41.6  & 50.1 \\ \cmidrule{2-6}
 & \multirow{2}{*}{1759h} & 30h & 40.7 & 32.5 & 31.2 \\
 &  & 433h & 38.6 & 28.6 & 26.0 \\
\bottomrule[0.1em]
\end{tabular}
}
 \label{main_results}
\end{table}

\begin{table*}[!htbp] 
    \centering
    \setlength{\tabcolsep}{10pt}

    \caption{\textbf{Performance comparison on LRS3 test set in low-resource setting.} In this setting, only 30h of LRS3 trainval set is utilized for finetuning after pretraining on unlabeled data from either LRS3 (433h) or LRS3+VoxCeleb2-en (1759h) data. `Base' and `Large' denote the size of the pretrained video encoder employed. Our Lip2Vec achieves favorable gains across different settings. Furthermore, compared to other approaches that require an auto-regressive decoder (CE), our inference speed is significantly higher due to CTC decoding. ${\dagger}$ denotes that our method does not utilize labeled video-text data during finetuning, but uses unlabeled video-audio pairs for the same.}
    \vspace{2mm}

    
     \adjustbox{width=0.75\textwidth}{
    \begin{tabular}
    {l l c c c c} 
     \toprule[0.1em]
      & \textbf{Method} & \textbf{Unlabeled AV data} & \textbf{Labeled Data}  & \textbf{Decoding} & \textbf{VSR} \\
     \toprule[0.1em]
 
     \multirow{10}{*}{\textbf{Base}} & AV-HuBERT~\cite{shi2022learning}& 433h & 30h & CTC & 55.3  \\ 
      & AV-HuBERT~\cite{shi2022learning}& 433h & 30h & CE & 51.8  \\ 
     & RAVen~\cite{haliassos2022jointly} & 433h & 30h & CTC+CE & 47.0 \\
     & VATLM~\cite{zhu2022vatlm} & 433h & 30h & CE & 48.0\\
     & \mycc \textbf{Ours: Lip2Vec} & \mycc 433h & \mycc 30h$^{\dagger}$ & \mycc CTC & \mycc 49.5 \\
     \cmidrule{2-6}
     & AV-HuBERT~\cite{shi2022learning,shi2022robust}& 1759h & 30h & CTC & 47.3 \\
     & AV-HuBERT~\cite{shi2022learning,shi2022robust}& 1759h & 30h& CE & 46.1 \\
     & RAVen~\cite{haliassos2022jointly} & 1759h & 30h& CTC+CE & 40.2 \\
     & VATLM~\cite{zhu2022vatlm} & 1759h & 30h& CE & 42.6\\
     & \mycc \textbf{Ours: Lip2Vec} & \mycc 1759h & \mycc 30h$^{\dagger}$ & \mycc CTC & \mycc 40.6\\
     \midrule
     \multirow{8}{*}{\textbf{Large}} & AV-HuBERT~\cite{shi2022learning}& 433h & 30h& CTC & 48.4 \\
      & AV-HuBERT~\cite{shi2022learning}& 433h & 30h& CE & 44.8 \\
     & \mycc \textbf{Ours: Lip2Vec} & \mycc 433h & \mycc 30h$^{\dagger}$ & \mycc CTC & \mycc 55.4\\
     \cmidrule{2-6}
     & AV-HuBERT~\cite{shi2022learning,shi2022robust}& 1759h & 30h & CTC & 40.7\\
     
     & AV-HuBERT~\cite{shi2022learning,shi2022robust}& 1759h & 30h & CE & 32.5\\
     & RAVen~\cite{haliassos2022jointly} & 1759h & 30h& CTC+CE & 33.1 \\
     & VATLM~\cite{zhu2022vatlm} & 1759h & 30h & CE & 31.6 \\
     & \mycc \textbf{Ours: Lip2Vec} & \mycc 1759h & \mycc 30h$^{\dagger}$ & \mycc CTC & \mycc 31.2\\
     \bottomrule[0.1em]
    \end{tabular}}

    \label{sota_low_res}
\end{table*}

\noindent\textbf{Finetuning \vs latent-to-latent:} \reftab{main_results} shows the performance comparison between supervised finetuning and our proposed latent-to-latent training in terms of WER score on the LRS3 test set. For both settings, identical pretrained video encoder from~\cite{shi2022learning} is utilized. The supervised finetuning using AV-HuBERT~\cite{shi2022learning} is performed with either a linear layer (CTC) or a decoder (CE) using labeled video-text pairs. In contrast, our latent-to-latent training employs unlabeled video-audio pair for training the prior network alone while the pretrained video encoder and ASR decoder (Wav2Vec2.0) are kept frozen. We observe that our latent-to-latent approach obtains consistent improvements across different settings. 
However, we observe that when the large video encoder is pretrained only on LRS3 (433h), the supervised finetuning achieves better performance. This is likely due to the large encoder overfitting to the pretraining data while being less generalizable and being prone to change at the finetuning stage to fit the labeled video-text pairs. Since the latent-to-latent procedure does not involve training the video encoder, our approach suffers when the pretrained video encoder is not generalizable. Such an issue does not arise for the base video encoder or when pretraining is performed on LRS3+VoxCeleb2-en (1759h), which helps in obtaining robust video representations that are better suited for latent-to-latent learning. 
It is also worth mentioning that Wav2Vec2.0 achieves 6.2 WER on the LRS3 test set. Furthermore, when using the large encoder pretrained on LRS3+VoxCeleb-En (1759h) and finetuning on 433h, our approach achieves the best WER score of $26.0$, with gains of $12.6$ and $2.6$ over the supervised CTC and CE finetuning, respectively. These results show the efficacy of our latent-to-latent learning approach for the VSR task.

\noindent\textbf{State-of-the-art comparison:} Here, we compare the Lip2Vec approach to SoTA VSR approaches on the LRS3 test set. Tables \ref{sota_low_res} and \ref{sota_high_res} show the performance comparison in terms of WER for the low-resource and high-resource settings, respectively. While the low-resource setting denotes that finetuning is performed with only 30h of LRS3 trainval data, the high-resource setting indicates finetuning with 433h of LRS3. Supervised methods using varying labeled data are also reported in \reftab{sota_high_res} for comparison. We observe that our Lip2Vec performs favorably against existing approaches across different settings. Furthermore, the approach depends on the generalizability of the pretrained video encoder representations. Because training the Lip2Vec does not utilize labeled video-text pairs in addition to freezing the parameters of the video encoder. This is in contrast to the supervised finetuning, which is likely to significantly vary the video encoder parameters to align for text decoding. Furthermore, from \reftab{sota_high_res}, we observe our Lip2Vec trained with large encoder with 1759h of pretraining to obtain the best results of $26.0$. This results in gains of $2.6$, $2.2$ and $2.4$ over AV-HuBERT, RAVen and VATLM, respectively, when self-training (\ie, pseudo-labeling the data and additionally using them for finetuning) is not employed by these approaches.





\begin{table*}[!htbp] 
    \centering
    \setlength{\tabcolsep}{10pt}

    \caption{\textbf{Performance comparison on LRS3 test set in high-resource setting.} `Base' and `Large' denote the size of the self-supervised video encoder employed. Performance of supervised approaches are also reported. ${\dagger}$ denotes that our Lip2Vec does not utilize labeled video-text data during finetuning, but uses unlabeled video-audio pairs for the same. Our approach achieves favorable gains across different settings with significantly higher inference speed due to CTC decoding, compared to other approaches that require an auto-regressive decoder (CE).  Particularly, when using the large encoder pretrained on 1759h, our approach achieves the best score of $26.0$ and is on par with $25.9$ of~\cite{serdyuk2021audio} that utilizes 90k hours of labeled data in a supervised setting.}
    \vspace{2mm}
    
    
     \adjustbox{width=0.85\textwidth}{
    \begin{tabular}
    {l l c c c c} 
     \toprule[0.1em]
      & \textbf{Method} & \textbf{Unlabeled AV data} & \textbf{Labeled Data}  & \textbf{Decoding} & \textbf{VSR} \\
     \toprule[0.1em]
    \multirow{6}{*}{\textbf{Supervised}} & Afouras \etal~\cite{afouras2018deep} & - & 1519h & CE & 58.9 \\
    & Shillingford \etal \cite{Shillingford2018Large-ScaleRecognition} & - & 3886h & CTC & 55.1 \\
    & Ma \etal \cite{ma2022visual}& - & 813h & CTC+CE & 34.7 \\
    & Makino \etal \cite{makino2019recurrent} & - & 31000h & Transducer & 33.6\\
    & Prajwal \etal \cite{Prajwal2021Sub-wordAttention} & - & 2676h & CE & 30.7 \\
    & Serdyuk \etal \cite{serdyuk2021audio} & - & 90000h & Transducer & 25.9 \\
    & Chang \etal \cite{chang2023conformers} & - & 100000h & Transducer & 12.8 \\
    
    \midrule
     \multirow{10}{*}{\textbf{Self-Supervised}} & AV-HuBERT~\cite{shi2022learning}& 433h & 433h & CTC & 49.3  \\ 
      & AV-HuBERT~\cite{shi2022learning}& 433h & 433h & CE & 44.0  \\ 
     \multicolumn{1}{c}{\multirow{8}{*}{\textbf{Base}}} & RAVen~\cite{haliassos2022jointly} & 433h & 433h & CTC+CE & 39.1 \\
     & \mycc \textbf{Ours: Lip2Vec} & \mycc 433h & \mycc 433h$^{\dagger}$ & \mycc CTC & \mycc  42.0 \\
     \cmidrule{2-6}
     & AV-HuBERT~\cite{shi2022learning,shi2022robust}& 1759h & 433h & CTC & 43.0 \\
     & AV-HuBERT~\cite{shi2022learning,shi2022robust}& 1759h & 433h& CE & 34.8 \\
     & RAVen~\cite{haliassos2022jointly} & 1759h & 433h& CTC+CE & 33.1 \\
     & VATLM~\cite{zhu2022vatlm} & 1759h & 433h& CE & 34.2 \\
     & \mycc \textbf{Ours: Lip2Vec} & \mycc 1759h & \mycc 433h$^{\dagger}$ & \mycc CTC & \mycc 34.1 \\
     \midrule
     \multirow{11}{*}{\textbf{Self-Supervised}} & AV-HuBERT~\cite{shi2022learning}& 433h & 433h& CTC & 44.3 \\
      & AV-HuBERT~\cite{shi2022learning}& 433h & 433h& CE & 41.6 \\
     \multicolumn{1}{c}{\multirow{9}{*}{\textbf{Large}}}  & \mycc \textbf{Ours: Lip2Vec} & \mycc 433h & \mycc 433h$^{\dagger}$ & \mycc CTC & \mycc 50.1 \\
     \cmidrule{2-6}
     & AV-HuBERT~\cite{shi2022learning,shi2022robust}& 1759h & 433h & CTC & 38.6\\
     & AV-HuBERT~\cite{shi2022learning,shi2022robust}& 1759h & 433h & CE & 28.6\\
     & AV-HuBERT~\cite{shi2022learning,shi2022robust} w/ self-training & 1759h & 433h & CE & 26.9\\
     & RAVen~\cite{haliassos2022jointly} & 1759h & 433h& CTC+CE & 28.2 \\
     & RAVen~\cite{haliassos2022jointly} w/ self-training  & 1759h & 433h& CTC+CE & 24.9 \\
     & VATLM~\cite{zhu2022vatlm} & 1759h & 433h & CE & 28.4 \\
     & VATLM~\cite{zhu2022vatlm} w/ self-training & 1759h & 433h & CE & 26.2 \\
     & \mycc \textbf{Ours: Lip2Vec} & \mycc 1759h & \mycc 433h$^{\dagger}$ & \mycc CTC & \mycc 26.0 \\
     \bottomrule[0.1em]
    \end{tabular}}

    \label{sota_high_res}
\end{table*}

\noindent\textbf{Results on VoxCeleb2-en:} In \reftab{vox_results}, we report the WER scores on three folds of the VoxCeleb2-en test set: the first fold is randomly selected $5$k videos, the second and third are subsets of this $5$k, where Wav2Vec2.0 obtains WER scores less than $30$ and $20$, respectively. We follow this procedure to reduce the bias and aim for a fair comparison as the labels are obtained with another ASR model (\ie, Whisper \cite{radford2022robust}). First, we observe that SoTA approaches fail to generalize under this benchmark. Both the model from \cite{ma2022visual} and the VTP~\cite{Prajwal2021Sub-wordAttention} scores are around 70 WER. It is worth mentioning that VTP was trained on a $2.7$k hours of video. As expected, the Wav2Vec2.0 gets relatively reasonable results (10 to 25 WER). Interestingly enough, our Lip2Vec approach tracks the Wav2Vec2.0 scores with an upper bound proportional to the quality of the prior network. When only trained on 30h of LRS3, our Lip2Vec deviates from Wav2Vec2.0 by an average WER score of 23 across the three folds, thereby showing the generalization capability of our approach. Note that self-trained variants of RAVen and AV-HuBERT are not considered for OOD generalization since they are trained on pseudo-labelled VoxCeleb2-en train set. It can be seen that our Lip2Vec also achieves consistent gains in terms of WER, in comparison to RAVen and AV-HuBERT across different folds, demonstrating better generalization to unseen or novel speakers. This trend holds for 433h finetuning

\begin{table}

\centering 
\caption{\textbf{Training the Lip2Vec on VoxCeleb2-en.} Comparing the effects of varying the training set on the WER scores on both LRS3 and VoxCeleb2-en test sets. We randomly select 30h from VoxCeleb2-en, and use it in different settings. We observe that the prior network can generalize to LRS3 when seing VoxCeleb2-en data only}

\setlength{\tabcolsep}{4pt}
\adjustbox{width=1\columnwidth}{
\begin{tabular}{l l | c c} 
\toprule[0.1em]
\multirow{2}{*}{\textbf{Encoder}}  & \multirow{2}{*}{\textbf{Training set }} & \multicolumn{2}{c}{\textbf{Test set}} \\
& & LRS3 & VoxCeleb2-en \\
 \toprule[0.1em]
\multirow{3}{*}{\textbf{Base}} & LRS3-30h &  40.6 &  58.2 \\
& LRS3+VoxCeleb2-en-60h & 40.1 & 54.6  \\
& VoxCeleb2-en-30h & 41.2 & 57.3 \\
 \midrule
\multirow{3}{*}{\textbf{Large}} & LRS3-30h & 31.2 & 39.4 \\
& LRS3+VoxCeleb2-en-60h & 30.4 & 33.1 \\
& VoxCeleb2-en-30h & 30.5 &  33.8 \\

\bottomrule[0.1em]
\vspace{-0.6cm}
\end{tabular}
}
 \label{vox_results_2}
\end{table}

 \textbf{Training on VoxCeleb2-en.} We investigate the impact of training with VoxCeleb2-en~\cite{chung2018voxceleb2} data. In practice, this scenario might arise if one has access to a dataset comprising unlabelled lip sequences. Our Lip2Vec framework is a suitable fit for this setting as it does not require labeled video-text pairs to learn the prior network. We take advantage of this property and train the model variants on a low-resources (30h) setting of the VoxCeleb2-en dataset. \reftab{vox_results_2} shows the WER scores following various training sets on both LRS3 and VoxCeleb2-en test sets. As expected, combining 30h from VoxCeleb2-en with the LRS3 low-resources setting improves the performance on the LRS3 test set as compared to training on 30h of LRS3 only ($30.5$ \vs $31.2$ for large and $40.1$ \vs $40.6$ for base). It is worth mentioning that training on 30h of VoxCeleb2-en achieves similar WER compared to using LRS3 low-resource dataset. This highlights the robustness of the proposed Lip2Vec approach and its considerable advantages over the supervised finetuning. 

The Whisper \cite{radford2022robust} pseudo-labelled VoxCeleb2-en test set turns out to be more challenging due to the high variety of speakers, vocabulary, \etc. The Lip2Vec variants scores on this benchmark are still far from their performance on the LRS3 test set. Future works will focus on the generalization aspects on both LRS3 and VoxCeleb2-en test sets.

\noindent\textbf{Inference speed:} As model efficiency is a key factor for real-world VSR applications, \reftab{vox_results} shows a GPU runtime comparison (processing time per 100 frames) of the different approaches on sample videos. Compared with other approaches, our model exhibits a remarkable improvement, being over $10\times$ faster than VTP, which is the fastest model among the tested models. This is explained by the fact that CTC decoding does not require any computationally expensive auto-regressive procedures, beam search, \etc.

\begin{table}

\caption[Caption for LOF]{\textbf{Out-of-distribution generalization on VoxCeleb2-en test set in terms of WER.} The folds are selected using Wav2Vec2.0 scores. Base and Large models are denoted by $^{*}$ and $^{\dagger}$. All models are fine-tuned on the 30h low-resource setting of LRS3. The performance on LRS3 test set is also shown for ease of reference. The last column reports the average computational load in seconds for decoding 100 frames video (4 seconds) on a single Nvidia A100.\vspace{0.1cm}}

\centering
\setlength{\tabcolsep}{9pt}
     \adjustbox{width=1.0\columnwidth}{

\begin{tabular}{l  ccccc}

\toprule[0.1em]
    \multirow{2}{*}{\textbf{Model}}& \multirow{2}{*}{Unlabeled} &\multicolumn{3}{c}{Video folds} & \multirow{2}{*}{Runtime (s)}  \\
    \cline{3-5}
     & &  01 & 02 & 03 &  \\
    \midrule
    
     Wav2Vec2.0 ~\cite{baevski2020wav2vec} & -- & 25.1 & 15.0 & 10.1 & 0.05 \\
     \midrule
    Ma et al.~\cite{ma2022visual}& -- & 69.4 & 64.1 & 61.3 &  3.91 \\
    VTP~\cite{prajwal2022sub}& -- & 71.7 & 69.1 & 66.9 & 0.97 \\

    \midrule

    RAVen$^{*}$ & 433h & 78.2  &  74.3  &  72.3 & --  \\
    
    AV-HuBERT$^{*}$ & 433h  &  79.1 &  76.3 &  73.2 & -- \\
    
    \mycc  \textbf{Ours: Lip2Vec}$^{*}$  & \mycc 433h & \mycc  71.2 & \mycc  65.7 & \mycc  57.3  & \mycc 0.07 \\
       \midrule[0.005em]
     
    RAVen$^{*}$ & 1759h  &  61.2 &  56.1 &  53.4 & -- \\
    AV-HuBERT$^{*}$ & 1759h &  71.1 &  66.2 &  62.9  & -- \\
    \mycc    \textbf{Ours: Lip2Vec}$^{*}$  & \mycc  1759h  &  \mycc  58.1 & \mycc  53.4 & \mycc  49.2 & \mycc 0.07  \\
    
    \midrule[0.1em]
    
     AV-HuBERT$^{\dagger}$ & 433h &  78.1 &  75.6 &  71.4 & -- \\
     \mycc  \textbf{Ours: Lip2Vec}$^{\dagger}$  & \mycc  433h    & \mycc  77.3  & \mycc  70.4 & \mycc  61.8 & \mycc 0.09  \\
       \midrule[0.005em]
      
    RAVen$^{\dagger}$ & 1759h   &  \textbf{47.5} &  42.8 &  39.4  & -- \\
     
    AV-HuBERT$^{\dagger}$ & 1759h &   49.7 &  44.6 &  41.2 & -- \\
    
    \mycc  \textbf{Ours: Lip2Vec}$^{\dagger}$  & \mycc  1759h  & \mycc  48.1 & \mycc  \textbf{39.4}  & \mycc  \textbf{33.2} & \mycc 0.09 \\
 
\bottomrule[0.1em]
\vspace{-0.5cm}
\end{tabular}
}
\label{vox_results}
\footnotetext{We select methods based on the source code availability on Github}
\end{table}

\subsection{Ablation Study}

Here, we evaluate the performance of our Lip2Vec when ablating the key components: varying the hyperparameter $\alpha$ for $\mathcal{L}_{mse}$ loss and the masking function $\mathcal{M}(\cdot)$. For this study, evaluation is conducted on LRS3 test set and the large video encoder (self-supervisedly pretrained on 1759 hours of LRS3 and VoxCeleb2-en) is employed.

\noindent\textbf{Impact of varying $\alpha$:} \reftab{tab:alpha_vary} presents the performance of our framework when the hyperparameter weight $\alpha$ (\refeq{eq3}) is varied. We observe that higher values of $\alpha$  degrade the performance since the similarity between predicted representations $\mathbf{z_{asr}^g}$ and target $\mathbf{z_{asr}}$ diverges due to the gradients from $\mathcal{L}_{mse}$ dominating over $\mathcal{L}_{cosine}$. Furthermore, when training for a fixed number of epochs, $\alpha{=}0$ achieves a WER score of 34.6 compared to the best results of 31.2 when $\alpha{=}0.01$. We also observe that training longer without MSE loss (denoted by $\dagger$ in \reftab{tab:alpha_vary}) can achieve a WER score of 31.4, indicating that $\mathcal{L}_{mse}$ aids in faster training convergence.

\begin{table}
\centering
\setlength{\tabcolsep}{8pt}
\caption{\textbf{Impact of varying $\alpha$.} WER comparison on LRS3 test set when varying the weight $\alpha$ for $\mathcal{L}_{mse}$. When $\alpha$ is increased beyond 0.05, the $\mathcal{L}_{mse}$ dominates over $\mathcal{L}_{cosine}$, resulting in $\mathbf{z_{asr}^g}$ diverging from the target $\mathbf{z_{asr}}$. While performance is slightly degraded without MSE loss for a fixed training budget, longer training (denoted by $\dagger$) can reach similar optimal performance as with $\alpha{=}0.01$, validating that $\mathcal{L}_{mse}$ improves the convergence. \vspace{0.1cm}\label{tab:alpha_vary}}

\begin{tabular}{cccccc}
\toprule
$\alpha$ &  0.0 & 0.0$^\dagger$ & 0.01 & 0.2 & 0.5  \\
\midrule
\textbf{WER} & 34.6 & 31.4 & 31.2 & 52.1 & 91.3 \\
\bottomrule
\vspace{-0.5cm}
\end{tabular}
\end{table}

\noindent\textbf{Impact of different masking strategies:} From \reftab{tab:mask_strategy}, we observe that not masking the audio representations $\mathbf{z_{asr}}$ results in the prior network learning a shortcut from its input to output while ignoring the video representations and thereby performing poorly at test time when no audio is available. Similarly, maintaining the same masking probability $p$ throughout the training results in the prior network expecting the masked audio to be present at test time as well for generating synthetic $\mathbf{z_{asr}^g}$ accurately. In contrast, initializing $p$ to a low value of $0.3$ and gradually increasing it to $1.0$ (simulating no $\mathbf{z_{asr}}$  input) by the end of training enables the prior network to learn better representations $\mathbf{z_{asr}^g}$ from input $\mathbf{z_{v}}$. Consequently, our progressive masking achieves a WER score of $31.2$, thereby validating its efficacy for training. Additional results are provided in the supplementary.

\begin{table}
\centering
\caption{\textbf{Masking strategy.} WER comparison on LRS3 test with different masking strategies $\mathcal{M}(\cdot)$ for the audio representations $\mathbf{z_{asr}}$. No masking performs poorly since the prior network discards the $\mathbf{z_{v}}$ input. Similarly, masking with same probability ($p$) throughout the training shows only marginal improvement over no masking. The best results are obtained with the proposed progressive masking, where $p$ is gradually increased during the training.\vspace{0.1cm}\label{tab:mask_strategy}}
\setlength{\tabcolsep}{8pt}
\begin{tabular}{lc}
\toprule
 \textbf{Masking} &  \textbf{WER}\\
\midrule
No Masking & 75.2 \\
50\% Masking & 66.9 \\
80\% Masking & 61.5 \\
100\% Masking & 65.3 \\
\textbf{Progressive Masking} & 31.2\\
\bottomrule
\vspace{-0.6cm}
\end{tabular}
\end{table}




\section{Discussion and Future Work}
\noindent \textbf{Supervised \vs self-supervised video encoder:} As discussed in the experiments, we employed a self-supervised video encoder from AV-HuBERT~\cite{shi2022learning} for training the prior network. In contrast, here, we  evaluate the efficacy of a supervised video encoder in the Lip2Vec framework by utilizing the encoder from \cite{ma2022visual}. For this experiment, we train the prior network following the low resources setting. This achieves WER scores of 45.0 and 76 on LRS3 and VoxCeleb2-en test sets, respectively. This is likely due to the explicit text supervision, which trains the video encoder to output representations aligned with the text decoding task rather than trained towards better representing the lip movements. This shows that self-supervised encoders are highly suited for learning the latent-to-latent mappings and are better generalizable. 
This is also supported by the findings in neuroscience research, which demonstrate that silent lip-reading signal first synthesizes a coarse-grained auditory speech representation in early auditory cortices. Then, the right angular gyrus excites the temporal visual speech area, extracts and possibly predicts the slower features of lip movements. Finally, the auditory cortices are fed with this signal to decode as an audio signal \cite{opoku2021visual, keitel2020shared}.
%
Consequently, our approach opens a new line of research for exploring the subtle definition of visual speech recognition embedded in the human brain \cite{bourguignon2020lip}. 

\noindent\textbf{VSR as interpolation \vs extrapolation:} Most perception problems are interpolative in their nature~\cite{chollet2019measure} and satisfy the manifold hypothesis~\cite{fefferman2016testing}. These tasks are intuitive for humans, and are usually solved in the early layers of the visual cortex in a matter of milliseconds (\ie, classification, recognition, \etc) \cite{kosaka2003neural, todorov2012role}. For such problems, deep learning is a perfect fit with its ability to perform non-linear interpolation in a complex high-dimensional manifold, enabling arbitrary complex behavior~\cite{summerfield2022natural,chollet2019measure}. However, lip-reading experts allude to a high-level step-wise and iterative reasoning to solve the task. This likely suggests that VSR has some higher level of extrapolation as compared to the common perception tasks. Thus, we hypothesize that learning the manifold transfer without exposing the lip sequences explicitly to the text labels would induce some interpolation, thereby allowing for a better generalization. We believe improving the prior network by leveraging  better training procedures and architectures such as \cite{Rombach2021High-ResolutionModels} would be an important future research direction for tightening up the bound with the ASR performance.

\noindent\textbf{Impact of fine-tuning on learned self-supervised encoder:} From our experiments above, we observed that supervised video encoders and models pretrained on LRS3 only, are not suitable for latent-to-latent learning. A potential future direction includes studying the effect of text labels on self-supervised learned weights using measures such as Center Kernel Alignment (CKA)~\cite{kornblith2019similarity} for obtaining deeper insights into the VSR task. 



\section{Conclusion}
We introduced Lip2Vec, a simple VSR framework that makes the most of ASR and VSR models by combining 
knowledge acquired by an off-the-shelf VSR encoder and an ASR model. The approach exploits the latent space structure to perform inter modality mapping, and learns how to transfer the visual representations to a suitable decoding space. Results on various benchmarks  demonstrated the competitiveness and robustness of the approach. We believe this is an important step towards better VSR modeling using latent-to-latent methods. In summary, the results and discussions presented in the paper along with those in the appendices demonstrate the efficacy of our Lip2Vec approach for the task of visual speech recognition.

{\small
\bibliographystyle{ieee_fullname}
\bibliography{egpaper_final.bib}
}

\clearpage

\appendix

\setcounter{table}{0}
\setcounter{figure}{0}
\renewcommand{\thetable}{A.\arabic{table}}
\renewcommand{\thefigure}{A.\arabic{figure}}

\section*{\LARGE Appendices}
We present additional quantitative and qualitative results of the Lip2Vec approach addressing the problem of visual speech recognition.  
\section{Varying the ASR Model and Video Encoder\label{sec:ext_models}}

The prior network $\boldsymbol{f_{\theta}}(\cdot)$ in our Lip2Vec framework can be potentially trained with different off-the-shelf (pretrained) ASR models and video encoders. 
Here, we evaluate the performance of our Lip2Vec approach when utilizing VQ-Wav2Vec~\cite{baevski2019vq} as ASR model and VATLM~\cite{zhu2022vatlm} as the video encoder.

\noindent\textbf{ASR model:} The choice of utilizing VQ-Wav2Vec as an alternate ASR model is motivated by the fact that it is semantically different from Wav2Vec2.0, as it relies on a discrete latent space. Particularly, the model first encodes an input audio signal as vector quantized (VQ) representations through a codebook learned on top of the feature extractor. Then, the resulting discrete representations of the audio are input to BERT \cite{devlin2018bert}, which outputs enhanced representations based on their respective surrounding context. Finally, an acoustic model is utilized to predict text from the BERT output representations. While pretrained VQ-Wav2Vec and BERT models are readily avialable\footnote{\url{https://github.com/facebookresearch/fairseq/blob/main/examples/wav2vec/README.md\#vq-wav2vec}}, the associated acoustic model is not. Therefore, we train a $6$-layer transformer decoder (CE auto-regressive decoding) along with a linear layer (for CTC decoding) on the BERT representations using the audio-text pairs in LRS3 training set. This acoustic model obtains 11.2 WER on the LRS3 test set when using CE+CTC decoding. 

Utilizing this VQ-Wav2Vec in our Lip2Vec indeed requires changing the prior network training objective to deal with codebook indices instead of continuous audio representations. Thus, we plug a classification head on the prior output to predict the codebook indices. Hence, we replace the cosine similarity loss with a standard cross entropy loss. \reftab{vq_results} shows the performance of our Lip2vec when using VQ-Wav2Vec as the ASR model in the low-resource setting (30h of finetuning data). We observe that it performs comparably with supervised finetuning of \cite{shi2022learning} across different settings, while requiring similar complexity due to CE+CTC decoding. The performance of our Lip2Vec when using Wav2Vec2.0 ASR model with CTC decoding alone is also shown for ease of comparison.

\begin{table}[t]

\centering 
\caption{\textbf{Supervised finetuning \vs latent-to-latent training.} Comparison in terms of WER on LRS3 test set is shown. The same pretrained video encoder from AV-HuBERT~\cite{shi2022learning} is finetuned (supervised w/ CE) or utilized for training the prior network in our Lip2Vec with two different ASR models: VQ-Wav2Vec and Wav2Vec2.0.}

\setlength{\tabcolsep}{3pt}
\adjustbox{width=1\columnwidth}{
\begin{tabular}{lcc|ccc}
\toprule[0.1em]
 \multirow{2}{*}{\textbf{Encoder}} & \multirow{2}{*}{\textbf{Pretrain}} & \multicolumn{1}{c|}{\multirow{2}{*}{\textbf{Finetune }}}& \textbf{Supervised} & \multicolumn{2}{c}{\textbf{Ours: Lip2Vec}} \\
 &  & \multicolumn{1}{c|}{} & S2S w/ CE & VQ-Wav2Vec  & Wav2Vec2.0     \\
 \toprule[0.1em]
\multirow{2}{*}{\textbf{Base}} & \multirow{1}{*}{433h} & 30h & 51.8 & 54.0 & 49.5 \\
 & \multirow{1}{*}{1759h} & 30h & 46.1 & 42.2 & 40.6 \\
 \midrule
\multirow{2}{*}{\textbf{Large}} & \multirow{1}{*}{433h} & 30h & 44.8 & 57.5 &  55.4\\
 & \multirow{1}{*}{1759h} & 30h & 32.5 & 33.5 & 31.2 \\
\bottomrule[0.1em]
\vspace{-0.6cm}
\end{tabular}
}
 \label{vq_results}
\end{table}

\begin{table}[t]

\centering 
\caption{\textbf{AV-HuBERT \vs VATLM as video encoder.} Comparison in terms of WER on LRS3 test set is shown. The pretrained video encoders from AV-HuBERT~\cite{shi2022learning} and VATLM~\cite{zhu2022vatlm} are utilized for training the prior network in our Lip2Vec framework. The same ASR model (Wav2Vec2.0) is utilized for both experiments.}

\setlength{\tabcolsep}{6pt}
\adjustbox{width=1\columnwidth}{
\begin{tabular}{lcc|cc}
\toprule[0.1em]
 \multirow{2}{*}{\textbf{Encoder}} & \multirow{2}{*}{\textbf{Pretrain}} & \multicolumn{1}{c|}{\multirow{2}{*}{\textbf{Finetune }}} & \multicolumn{2}{c}{\textbf{Video Encoder}} \\
 &  & \multicolumn{1}{c|}{} & VATLM & AV-HuBERT    \\
 \toprule[0.1em]
 \multirow{1}{*}{\textbf{Base}} & \multirow{1}{*}{1759h} & 30h  & 42.5 & 40.6 \\

  \midrule
  
\multirow{1}{*}{\textbf{Large}} & \multirow{1}{*}{1759h} & 30h & 33.0 &  31.2\\
\bottomrule[0.1em]
\end{tabular}
}
 \label{vatlm_results}
\end{table}

\noindent\textbf{Video encoder:} Here, we evaluate the performance of Lip2Vec when utilizing a different self-supervised video encoder from VATLM~\cite{zhu2022vatlm}. It is worth mentioning that VATLM follows the same architecture and training procedure as AV-HuBERT. However, VATLM additionally utilizes the text modality during pretraining to enhance the features and promote for a unified latent space. 
\reftab{vatlm_results} shows the performance comparison when utilizing AV-HuBERT and VATLM encoders for training our prior network in the low-resource setting. Both encoders are pretrained on 1759h of LRS3+VoxCeleb2-en data. 

Since VATLM utilizes text modality during pretraining, the resulting encoder representations are likely to be better aligned to the task of text prediction than for representing the lip sequences. Despite this, the VATLM encoder-based Lip2Vec achieves WER scores of $42.5$ and $33.0$ WER when using the Base and Large encoder architectures, respectively and performs comparably with the AV-HuBERT encoder-based Lip2Vec.

In summary, the aforementioned results and discussion demonstrate the capability of our Lip2Vec approach to successfully adapt to different ASR models and video encoders for learning the prior network using unlabelled video-audio pairs. Consequently, the Lip2Vec forms a viable alternative to video-text supervised finetuning.

\section{Additional Results\label{sec:vid_len_poses}}
In this section, we analyse the robustness of our Lip2Vec approach when varying the video sequence lengths and head poses of the speaker at test time. This is followed by a discussion on common failure cases and model consistency. 


\begin{table}
\centering
\caption{\textbf{Impact of varying video length.} Comparison is shown in terms of WER on the LRS3 test set (denoted by All) along with four subsets of the same test set partitioned based on the length of the videos. LR and HR denote the low- and high-resource training with 30h and 433h of LRS3, respectively. Typically, text prediction is degraded for short sequences (less than 2 seconds) due to lack of contextual information during visual feature encoding.}
\setlength{\tabcolsep}{5pt}
\adjustbox{width=1\columnwidth}{
\begin{tabular}{l c | c c c c}
\toprule
 \multirow{2}{*}{\textbf{Model}} & \multirow{2}{*}{\textbf{All}} & \multicolumn{4}{c}{\textbf{Video Length (in seconds)}} \\
  &  & 0-2 & {2-4} & {4-6} & {$>$ 6} \\
\midrule
VTP~\cite{prajwal2022sub} & 40.6 & 46.2 & 41.5 & 36.8 & 29.4 \\

VTP~\cite{prajwal2022sub} (2676h) & 30.7 & 38.0 & 31.1 & 24.5 & 21.3 \\

Ma \etal~\cite{ma2022visual} & 32.3 & 41.1 & 31.6 & 22.5 & 17.1 \\

\textbf{Ours: Lip2Vec} (LR) & 31.2 & 38.8 & 31.7 & 22.7 & 17.2 \\

\textbf{Ours: Lip2Vec} (HR) & 26.0 & 34.2 & 24.5 & 15.9 & 17.2 \\

\bottomrule
\vspace{-0.6cm}
\end{tabular}
}
\label{len_test}
\end{table}

\noindent\textbf{Varying the Video Length:} \reftab{len_test} shows the performance comparison on different folds obtained by partitioning the LRS3 test set based on the video sequence length. We observe that shorter videos (less than $2$ seconds, \ie, $50$ frames) present a bottleneck, which results in performance degradation of the approaches from their corresponding average WER on the whole LRS3 test set (denoted as All in \reftab{len_test}). This is likely due to the lack of rich contextual features in shorter video sequences, which leads to sub-optimal temporal modeling in the video encoder. Consequently, the resulting representations output by the video encoder are not sufficiently discriminative for decoding the text correctly. Furthermore, we observe that the SoTA approaches and our Lip2Vec generally perform better with longer videos as input, indicating the importance of temporal modeling of visual features for accurate text decoding. However, targeting this issue is an important line of research to follow.

\begin{figure*}
\makebox[\linewidth]{
    \centering
    \includegraphics[width=1.00\linewidth]{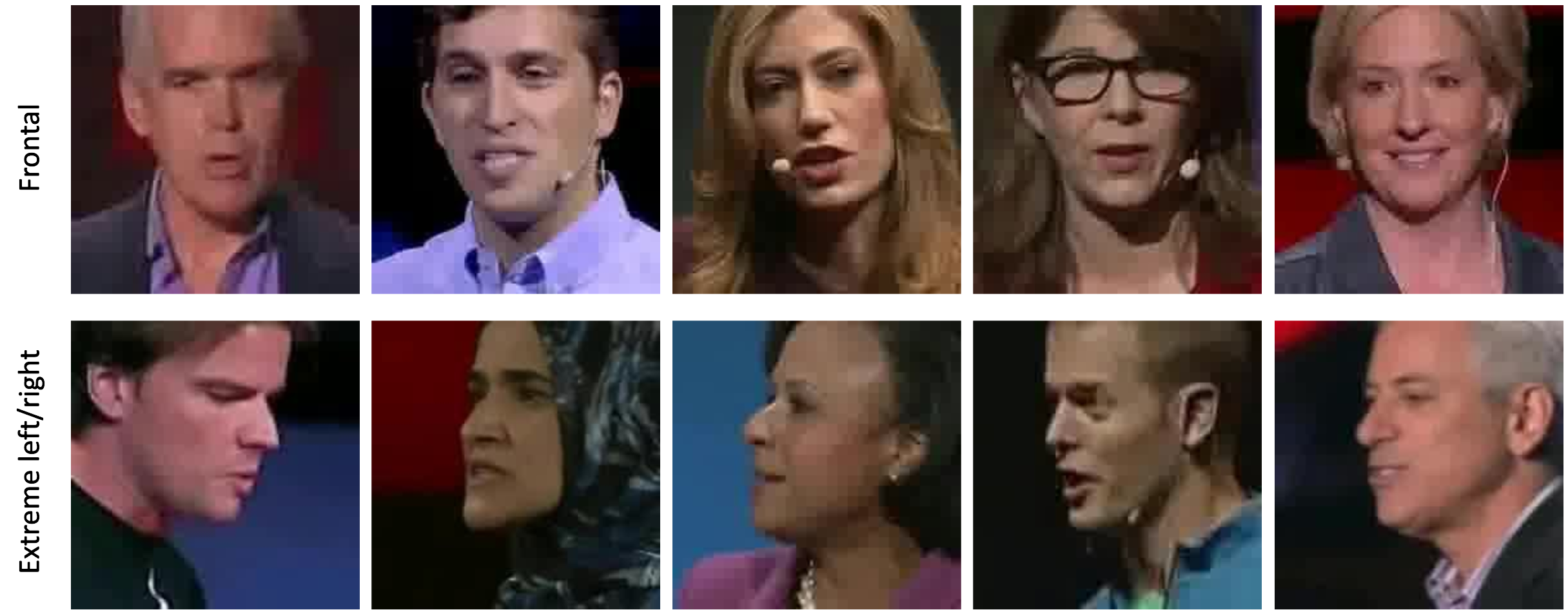}}
    \caption{\textbf{Frontal \vs extreme head poses in videos.} Top and bottom rows show example frames from videos having speakers with frontal and extreme (right/left) head poses, respectively. The lips sequences in extreme head poses are not completely visible and are likely to result in less discriminative representations output by the video encoders.}

    \label{poses}
\end{figure*}

\noindent\textbf{Varying Head Poses:} \reffig{poses} shows example frames from videos with frontal and extreme head poses in the LRS3 dataset. For this experiment, we select random $132$ videos from LRS3 test for each of the subsets: frontal and extreme.
We recover the 3D head pose by using a recently introduced method \cite{Filntisis2022VisualVideos} targeting monocular 3D face reconstruction from talking face videos. Given a parametric 3D model \cite{Li2017LearningScans} built from large datasets of 3D scans of human
faces, this approach regresses the 3D model parameters that best fit to each image frame. We consider frontal and extreme based on predefined face angles.
\reftab{head_pose} shows the performance comparison between different approaches on both theses subsets, in terms of WER. We observe that decoding text from videos with extreme head poses is challenging since the lip sequences in such videos are only partially visible, resulting in less discriminative representations output by the video encoder. Among the approaches, only VTP achieves comparable results for both subsets. This is likely due to VTP utilizing the sequence of full images as input instead of the cropped lip sequences. 

In summary, the presented Lip2Vec framework that learns a prior network using video-audio pair data performs favorably in comparison to other approaches across different settings with varying video lengths and head poses.

\begin{table}
\centering
\caption{\textbf{Impact of head pose.} Comparison is shown in terms of WER on the LRS3 test set (denoted by All) along with two subsets: Frontal and Extreme, partitioned based on the head pose of the speaker in the video. LR and HR denote the low- and high-resource training with 30h and 433h of LRS3, respectively. Decoding text from partial/occluded lip motion at extreme head poses is challenging compared to frontal videos, where the lips are fully visible. See text for more details.}
\setlength{\tabcolsep}{10pt}
\adjustbox{width=\columnwidth}{
\begin{tabular}{l c c c }
\toprule
 \textbf{Model} & \textbf{All} & \textbf{Frontal} & \textbf{Extreme} \\
\midrule
VTP~\cite{prajwal2022sub} & 40.6 & 38.5 & 37.7 \\

VTP~\cite{prajwal2022sub} (2676h) & 30.7 & 29.4 & 28.4 \\

Ma \etal~\cite{ma2022visual} & 32.3 & 28.8 & 33.4 \\

\textbf{Ours: Lip2Vec} (LR) & 31.2 & 25.9 & 33.4 \\

\textbf{Ours: Lip2Vec} (HR) & 26.0 & 19.4 & 29.4 \\

\bottomrule
\vspace{-0.6cm}
\end{tabular}
}
\label{head_pose}
\end{table}

\begin{figure*}
\makebox[\linewidth]{
    \centering
    \includegraphics[width=1.00\linewidth]{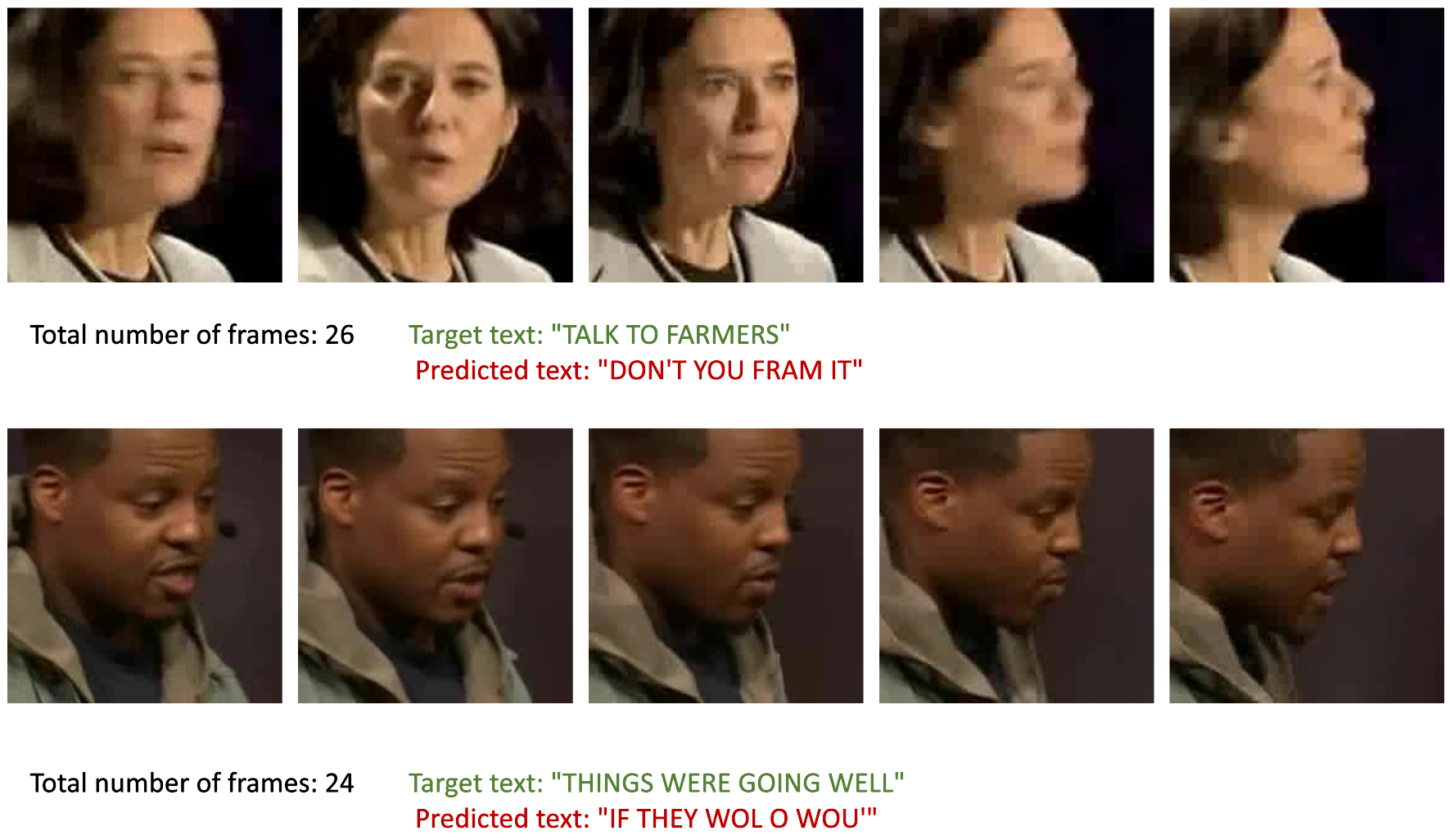}}

    \caption{\textbf{Illustration of failure cases.} We observe the text decoding to be less accurate in case of short videos (around 1 second), where contextual representation is difficult. Furthermore, rapid variation of poses with blurry frames (top row) and extreme poses (bottom row) present a challenge for accurate text decoding. It is worth mentioning that although the predicted sentence for the top row video is not accurate, it has the same lip motion as the target sentence (\ie, they are homophemes). }
    \label{failure}
\end{figure*}

\noindent\textbf{Failure cases:} \reffig{failure} illustrates example failure cases of the Lip2Vec framework. In the top row, the model fails to adapt to rapid head motion (the speaker turns the head suddenly from left to right while talking) in a short sequence. Additionally, the frames appear blurred due to the rapid motion, which likely affects the visual representations as well. The predicted sentence in this case, although incorrect, is still a homopheme and has the same lip motion as the target text. The bottom row example appears to be more challenging, since the subject has an extreme head pose all along the short sequence, leading to a set of poor visual representations and hence, failed decoding. A potential future direction, beyond the scope of the current work, could be to employ head pose normalization techniques as a preprocessing step to frontalize the videos and use them as input.

\begin{table}[t]
\centering
\caption{\textbf{Model consistency.} Performance comparison on LRS3 test set in terms of weighted mean ($\mu_{wer}$), standard deviation ($\sigma_{wer}$) and rank metric ($\mu_{wer}(1+\sigma_{wer})$). Our Lip2Vec achieves lower rank metric indicating the consistency of predictions.}
\setlength{\tabcolsep}{8pt}

\adjustbox{width=\columnwidth}{
\begin{tabular}{l c c c }
\toprule
 \textbf{Model} & $100 \times \mu_{wer}$ & $\sigma_{wer}$ &  $100 \times \mu_{wer}(1+\sigma_{wer})$ \\
\midrule
VTP~\cite{prajwal2022sub} & 30.7 & 0.38 & 42.4 \\

Ma~\cite{ma2022visual} & 32.5 & 0.42 & 46.2 \\

\textbf{Ours: Lip2Vec} (LR) & 31.2 & 0.30 & 40.6 \\

\textbf{Ours: Lip2Vec} (HR) & 26.0 & 0.29 & 33.5 \\

\bottomrule
\vspace{-0.6cm}
\end{tabular}
}
\label{stats_test}
\end{table}

\noindent\textbf{Model consistency:} The WER \cite{klakow2002testing} is the metric used for comparing different VSR models. However, given that the test set videos have varying target lengths, weighted average WER ($\mu_{wer}$) across the test set might not be sufficient for comparing different approaches, \eg, a model might fit precisely to some samples while having poor predictions for others. Furthermore, we observe that the WER distribution on the LRS3 test set is non-symmetric with more mass around 0-20, while the weighted standard deviation ($\sigma_{wer}$) is in the order of the mean. Thus, we combine both mean and standard deviation in a unified rank metric, as $\mu_{wer}(1+\sigma_{wer})$, to compare the models. Such a metric correctly penalizes models that achieve lower $\mu_{wer}$ at the cost of higher $\sigma_{wer}$. \reftab{stats_test} shows the performance comparison between our Lip2Vec and other approaches on LRS3 test set. We observe that our Lip2Vec achieves better results (lower is better), demonstrating the consistency in predicitions. In fact, our Lip2Vec (LR) has a higher $\mu_{wer}$ than VTP ($31.2$ \vs $30.7$) but achieves lower $\sigma_{wer}$. As a result, the final rank metric is better for our Lip2Vec ($40.6$ \vs $42.4$).

\end{document}